\documentclass[aps,twocolumn,prl,showpacs,superscriptaddress,groupedaddress]{revtex4}  
\usepackage{dcolumn}
\usepackage{bm}
\usepackage{color}

\def\ltape{\hbox{\ $<$\hskip -8pt\raise -4pt\hbox{$\sim$}\ }}
\def\gtape{\hbox{\ $>$\hskip -8pt\raise -4pt\hbox{$\sim$}\ }}

   \ifx\pdfoutput\undefined
   \usepackage[dvips]{graphicx}
   \else
   \usepackage[pdftex]{graphicx}
   \fi
   \usepackage{epstopdf}
   \DeclareGraphicsExtensions{.pdf,.gif,.jpg,.eps,.png}
  
   \usepackage{amsmath}
  \usepackage{amssymb}

\pdfoutput=1

\begin{document}

\title{Strongly localized magnetic reconnection by the super-Alfv\'enic shear flow}

\author{Yi-Hsin~Liu}
\affiliation{Dartmouth College, Hanover, NH 03750}
\author{M.~Hesse}
\affiliation{University of Bergen, Bergen, Norway}
\affiliation{Southwest Research Institute, San Antonio, TX 78238}
\author{F.~Guo}
\affiliation{Los Alamos National Laboratory, Los Alamos, NM 87545}
\author{H.~Li}
\affiliation{Los Alamos National Laboratory, Los Alamos, NM 87545}
\author{T. K. M.~Nakamura}
\affiliation{Space Research Institute, Austrian Academy of Sciences, Graz 8010, Austria}

\begin{abstract}

We demonstrate the dragging of the magnetic field by the super-Alfv\'enic shear flows out of the reconnection plane can strongly localize the reconnection x-line in collisionless plasmas, reversing the current direction at the x-line. Reconnection with this new morphology, which is impossible in resistive-magnetohydrodynamic (MHD), is enabled by electron inertia. 
Surprisingly, the quasi-steady reconnection rate remains of order $0.1$ even though the aspect ratio of the local x-line geometry is larger than unity. We explain this by examining the transport of the reconnected magnetic flux and the opening angle made by the upstream magnetic field, concluding that the reconnection rate is still limited by the constraint imposed at the inflow region. This study further suggests the nearly universal fast rate value of order $0.1$ cannot be explained by the physics of tearing modes, nor can it be explained by a universal localization mechanism.

\end{abstract}


\maketitle

{\it Introduction--}
Magnetic reconnection is a fundamental mechanism that converts magnetic energy into plasma kinetic energy by altering the connectivity of magnetic field lines \cite{zweibel09a}. While most studies focus on cases that do not have pre-existing flows upstream of the reconnection layer, reconnection can occur at plasma boundary layers where the flows on two sides of the current sheet are very different  \cite{TKMNakamura17a,eriksson16a,WLi16a, XMa14a, TKMNakamura13a, karimabadi13c, knoll02a,nykyri01a,fairfield00a,otto00a,QChen97a,belle-hamer95a,pu90a,cassak11a}. At the flank of Earth's magnetopause, it is established that the shear flow parallel to the anti-parallel magnetic fields can induce reconnection during the nonlinear development of Kelvin-Helmholtz vortices \cite{TKMNakamura17a,eriksson16a,WLi16a, XMa14a, TKMNakamura13a, karimabadi13c, knoll02a,nykyri01a,fairfield00a,otto00a,QChen97a,belle-hamer95a,pu90a}. On the other hand, the shear flow perpendicular to the anti-parallel magnetic field can also be significant \cite{hwang11a,XMa14a, XMa16a}, but its effect on reconnection is less clear. A similar situation also applies to relativistic jets that power gamma-ray bursts and active galactic nuclei blazars \cite{bromberg16a, giroletti04a}. The shear flows are super-Alfv\'enic and perpendicular to the helical magnetic fields inside the jet. Reconnection can take place in between these helical magnetic fields  \cite{bromberg16a}. A recent study showed that the field-line dragging effect by the shear flow out of the reconnection plane can significantly change the reconnection outflow structure \cite{XMa16a}, but its effect on the reconnection diffusion region physics remains unexplored.

In this Letter, we use kinetic simulations to study the reconnection diffusion region in the presence of an out-of-plane super-Alfv\'enic shear flow in pair plasmas. While the choice of parameters is not intended to address a particular observation, the simulation designed here serves as a proof-of-principle experiment that sheds new light on the reconnection rate problem. 
Surprisingly, a strongly localized x-line geometry is achieved by the field-line dragging effect of the flow shear, which induces an embedded sheet where the electric current reverses its direction.
This feature alters how the {\it frozen-in} condition is broken and how the magnetic energy is converted, compared to the standard regime [e.g.,\cite{goldman15a}]. It also illuminates the relationship between the {\it localization} of the diffusion region and the fast reconnection rate of order 0.1 \cite{parker73a,shay99a,birn01a};  
for decades, the morphological difference between the slow Sweet-Parker solution \cite{sweet58a,parker57a} and the fast Petschek solution \cite{petschek64a} had prompted the search for a ``universal''  localization mechanism, that leads to a short diffusion region and the fast rate of order $0.1$ \cite{parker73a,shay99a,birn01a}.
Dispersive-waves arising from the Hall effect were argued to provide the localization in collisionless plasmas \cite{birn01a,rogers01a,drake08a}, but the same rate is found in dispersive-less regimes \cite{bessho05a,daughton07a,swisdak08a,yhliu14a,tenbarge14a}. 
Recent progress in high Lundquist number MHD theory and simulations \cite{loureiro07a,samtaney09a,biskamp86a} demonstrates that fast-growing secondary tearing modes generate multiple x-lines that chop the long (Sweet-Parker) current sheet into shorter segments, resulting in a rate faster than the Sweet-Parter scaling \cite{YMHuang10a,uzdensky10a}. The same current filamentation tendency may also limit the current sheet extension of a single x-line in the collisionless limit, as implied by the sporadic generation of secondary plasmoids \cite{yhliu14a, daughton06a}.
These foster a popular conjecture that the fast rate of order 0.1 might be the result of localization universally provided by the tearing physics. For instance, one may argue that a single x-line is marginally stable to secondary tearing modes so that the aspect ratio of the diffusion region is subject to the marginally stable condition, $k\delta \gtrsim 1$. Here $k$ is the wavenumber of a tearing mode and $\delta$ is the half-thickness of the current sheet. This condition then implies a critical aspect ratio $\delta/(2\pi/k) = 1/2\pi \simeq 0.16$ for the diffusion region, and this seems to explain the fast rate. However, the rate discussed in this case remains of order 0.1 even when the localization mechanism is distinctly different, and the filamentation tendency within the reversed current can not regulate the length of the diffusion region.
These findings suggest that the fast rate of order 0.1 observed in disparate systems can not be explained by the tearing physics, nor by a ``universal'' localization mechanism. 
Instead, the same fast rate in this case can still be explained by the upper bound value imposed at the inflow region \cite{yhliu17a,cassak17a,yhliu18a}.

\begin{figure}
\includegraphics[width=8cm]{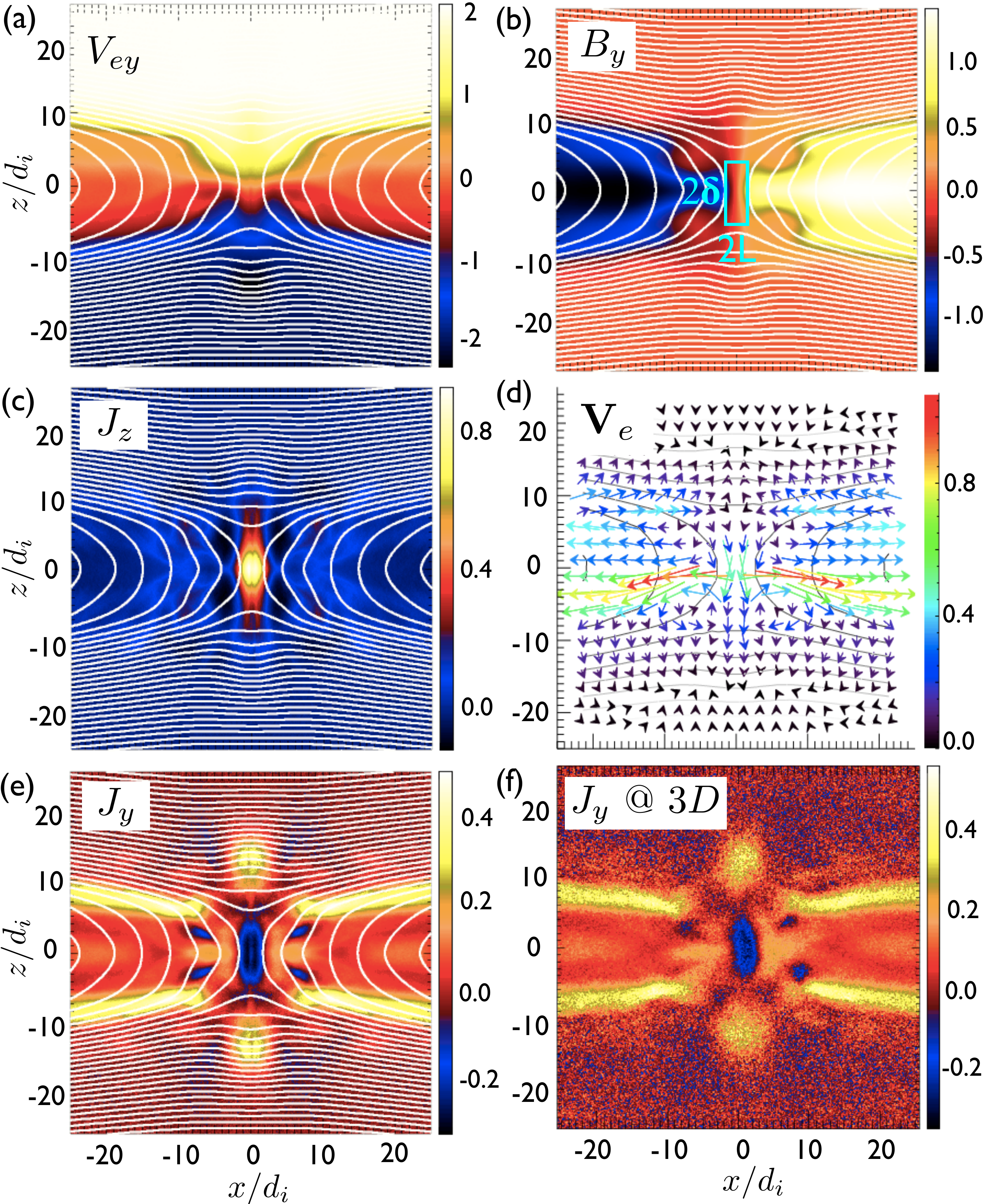} 
\caption {Quantities at time $120/\Omega_{ci}$. (a) $V_{ey}$ and the contour of the in-plane magnetic flux. (b) $B_y$. (c) $J_z$. (d) Vector plot of the in-plane electron flow. The color represents $|V_{e,xz}|\equiv (V_{ex}^2+V_{ez}^2)^{1/2}$. (e) $J_y$. (f) $J_y$ in a companion 3D simulation.
}
\label{morphology}
\end{figure}
{\it Simulation setup--}
Electron-positron plasmas with the mass ratio $m_i/m_e=1$ provide the simplest test bed for our study due to the mass symmetry, which excluded the Hall effect \cite{bessho05a}. This choice is also motivated by astrophysical applications [i.e., \cite{FGuo15a,FGuo14a,sironi14a,werner16a,yyuan16a,lyutikov17a}]. Simulations were performed using the particle-in-cell code {\it VPIC} \citep{bowers09a}, which solves the interaction between charged particles and electromagnetic fields. The initial magnetic profile is ${\bf B}=B_0 \mbox{tanh}(z/\lambda)\hat{x}$ and the initial $J_y >0$. The density is $n_j=n_0\mbox{sech}^2(z/\lambda)+n_b+\delta n_j$ where $j=i,e$ denotes ions (positrons) and electrons. An initial out-of-plane shear flow ${\bf V}=V_{shear} \mbox{tanh}(z/\lambda)\hat{y}$ is implemented, which produces an electric field $E_z=V_yB_x/c$. A self-consistent charge separation $\rho_c=e(\delta n_i-\delta n_e)=\partial_z E_z$ is calculated to satisfy the Poisson equation.  We assume $T_i/T_e=1$ and $\delta n_i=-\delta n_e$ to further ensure the symmetry of electron and positron motions. Pressure balance requires a uniform $P_i+P_e+B_x^2/8\pi-E_z^2/8\pi$, which determines the thermal speed $v_{th}/c$. Force balance of each species is also satisfied, which determines the drift speed of current carriers inside the current sheet. 

Densities are normalized to $n_0$. Spatial scales are normalized to the ion inertial length $d_i\equiv c/\omega_{pi}$, with ion plasma frequency $\omega_{pi}\equiv(4\pi n_0 e^2/m_i)^{1/2}$. Time scales are normalized to the ion gyro-frequency $\Omega_{ci}\equiv eB_0/m_i c$. In this simulation, $\omega_{pi}/\Omega_{ci}=4$ and $n_b=n_0$. The upstream Alfv\'en speed of pair plasma is $V_{A0}= B_0/\sqrt{8\pi n_0 m_i }\sim 0.25/\sqrt{2}\simeq 0.177c$ and $v_{th}/c\simeq 0.195$. 
Velocities, magnetic and electric fields are normalized to $V_{A0}$, $B_0$ and $B_0 V_{A0}/c$, respectively. The initial current sheet thickness is $\lambda=d_i$. The system size is $L_x \times L_z = 128 d_i \times 128 d_i$ with $2048 \times 2048$ grid points and $2000$ particles per cell. In the 3D case the system size is $L_x \times L_y \times L_z = 128 d_i \times 64 d_i \times 64 d_i$ with $2048 \times 1024 \times 1024$ grid points and $100$ particles per cell. The boundary conditions are periodic both in the $x$- and $y$-directions (i.e., 3D case), while in the $z$-direction they are conducting for fields and reflecting for particles. 

\begin{figure}
\includegraphics[width=6.5cm]{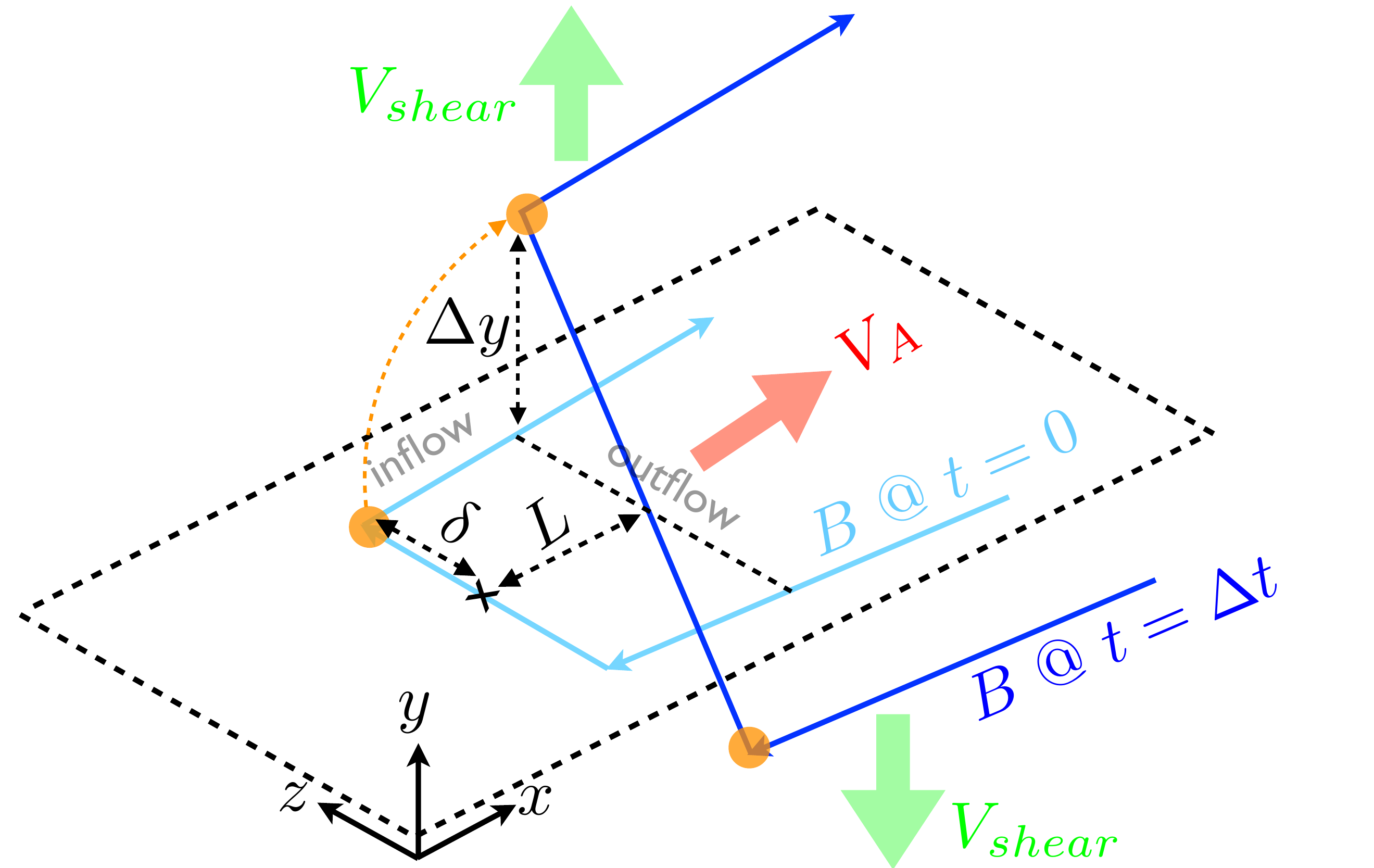} 
\caption {Dragging of the reconnected field by the shear flow.
}
\label{model}
\end{figure}

{ \it Formation of an embedded sheet with negative current density--} 
The morphology near the reconnection x-line with $V_{shear}=2 V_{A0}$ is shown in Fig.~\ref{morphology}. 
The dragging of the reconnected field $B_z$ by the out-of-plane shear flow $V_y$ (Fig.~\ref{morphology}(a)) generates a strong out-of-plane field $B_y$ (\ref{morphology}(b))\cite{XMa16a}. This $B_y$ of opposite sign sandwiches the x-line, driving a narrow current channel $J_z$ (\ref{morphology}(c)) that consists of high speed electrons streaming in the negative z-direction (\ref{morphology}(d)) and high speed positrons streaming in the positive z-direction. 
The strength of $B_y$ can be estimated by considering the local x-line geometry of dimension $2L \times 2\delta$ marked in Fig.~\ref{morphology}(b). The time for
the reconnected flux to be convected a distance $L$ by the Alfv\'enic outflow is $\Delta t \sim L/V_{A0}$. Meanwhile, the shear flow displaces the leg of the reconnected flux tube by $\Delta y \sim V_{shear} \Delta t$, as illustrated in Fig.~\ref{model}. The straightened reconnected field line at $x=L$ suggests $B_{y,out}/B_{z,out}\sim \Delta y/\delta$. Here ``out'' and ``in'' indicate the outflow and inflow regions. We also know $B_{z,out}/B_{x,in}\sim \delta/L$ from $\nabla\cdot {\bf B}=0$. Thus, $B_{y,out}/B_{x,in}\sim \Delta y/L\sim V_{shear}/V_{A0}$. When the shear flow is super-Alfv\'enic, magnetic pressure $B_{y,out}^2/8\pi$ becomes larger than $B_{x,in}^2/8\pi$. This difference will squeeze the x-line (where the initial thermal pressure was of order $B_{x,in}^2/8\pi$) in the x-direction.
The x-line is thus strongly localized. For a similar reason, the outflow region will also expand outwardly in the z-direction. 
Ampere's law further suggests 
\begin{equation}
J_y\simeq \frac{c}{4\pi}\left(\frac{B_{x,in}}{\delta}-\frac{B_{z,out}}{L}\right) \simeq \frac{cB_{x,in}}{4\pi\delta}\left(1-\frac{\delta^2}{L^2}\right). 
\label{Jy}
\end{equation}
When $\delta/L >1$, the current density with an opposite sign ($J_y < 0$) develops at this strongly localized x-line (Fig.~\ref{morphology}(e)).

It is important to note that secondary tearing modes do not form; they are not favored because the negative current density can only lead to a current filamentation that reverses the primary reconnection process. This fact has an important implication that will be discussed later. To demonstrate the robustness of such an embedded current sheet in 3D, a companion 3D run with $L_y =64 d_i$ is shown in Fig.~\ref{morphology}(f). Here we study how reconnection can proceed with this abnormal geometry.

\begin{figure}
\includegraphics[width=8cm]{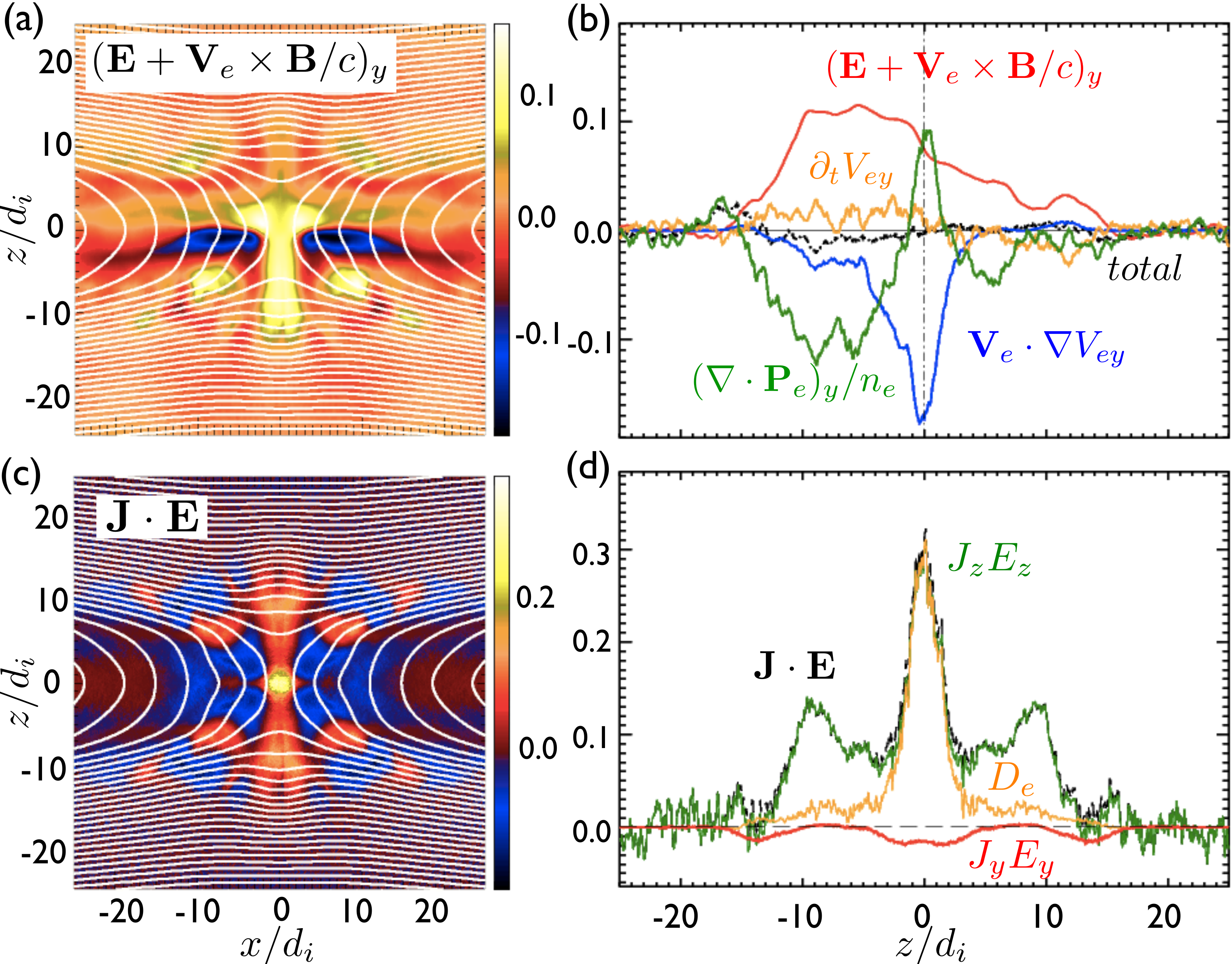} 
\caption {Quantities at time $120/\Omega_{ci}$. (a) Non-ideal electric field $({\bf E}+{\bf V}_e\times {\bf B}/c)_y$ and the contour of in-plane magnetic flux. (b) Composition of the non-ideal electric field along $x=0$. (c) Energy conversion measure ${\bf J}\cdot{\bf E}$. (d) Composition of the energy conversion measure ${\bf J}\cdot{\bf E}$ and the dissipation measure $D_e$ along $x=0$.
}
\label{dissipation_diffusion}
\end{figure}

{\it Frozen-in violation and energy conversion--} 
The formation of such an embedded sheet with a negative current density during reconnection is impossible in resistive-MHD because the negative $J_y$ and a positive  resistivity $\eta$ will make $E_y=\eta J_y < 0$ at the x-line and reverse the reconnection process. To show how this works in collisionless plasmas, we analyze the electron momentum equation, ${\bf E}_e'\equiv {\bf E}+{\bf V}_e\times {\bf B}/c =-\nabla\cdot {\bf P}/en_e-(m_e/e)({\bf V}_e\cdot\nabla {\bf V}_e+\partial_t {\bf V}_e)$, in the out-of-plane direction. A finite non-ideal electric field $E_{ey}'$ indicates the violation of the frozen-in condition for electrons. As shown in Fig.~\ref{dissipation_diffusion} (a), $E_{ey}'$ is still positive around the x-line, consistent with the reconnection flow pattern. The complex pattern of $E_{ey}'$ is asymmetric in the inflow direction. A similar observation applies to $E'_{iy}$ for positrons, which is a mirror reflection of the $E'_{ey}$ pattern with respect to the $z=0$ axis. Cuts along the $x=0$ axis in Fig.~\ref{dissipation_diffusion} (b) show that the dominant term that breaks the frozen-in condition at the x-line is electron inertia, ${\bf V}_e\cdot \nabla V_{ey}$. This is very different from a typical symmetric reconnection without shear flows, in which the $(\nabla\cdot {\bf P}_e)_y$ dominates at the x-line because the in-plane flow vanishes at the x-line [e.g.,\cite{hesse99a}]. The difference comes from the finite $V_{ez}$ at the x-line (as shown in Fig.~\ref{morphology}(d)), which contributes significantly through ${\bf V}_e\cdot \nabla V_{ey}\approx V_{ez}\partial V_{ey}/\partial z$ at the x-line. A similar observation is found in asymmetric reconnection \cite{hesse14a}. While $(\nabla\cdot {\bf P}_e)_y $ suppresses the reconnection electric field at the x-line, it contributes positively at other regions.

At first glance, a negative $J_y E_y$ seems to pose a problem in the energy conversion process of reconnection. Applying the Poynting theorem in a steady state, $\nabla\cdot {\bf S}=-{\bf J}\cdot {\bf E}$ where ${\bf S}={\bf E} \times {\bf B}/4\pi$ is the Poynting flux. A finite positive ${\bf J}\cdot {\bf E}$ inside the diffusion region suggests an energy conversion from the in-flowing reconnecting magnetic field to outflowing plasma kinetic energy \cite{birn10b}. Although the reconnecting component $J_y E_y$ is negative because $J_y< 0$, the total ${\bf J}\cdot {\bf E}$ surrounding the x-line is still positive as shown in Fig.~\ref{dissipation_diffusion}(c).  The dominant contribution of ${\bf J}\cdot {\bf E}$ in this case is the positive $J_z E_z$, as depicted by the green curve in Fig.~\ref{dissipation_diffusion}(d). Therefore, the magnetic energy is still converted to plasma energy even though it is not through the reconnection electric field $E_y$, as is typical [e.g.,\cite{goldman15a}]. A frame-independent measure $D_e$($\simeq {\bf J}\cdot{\bf E}'$) was used to quantify the dissipation \cite{zenitani11c}. It highly concentrates at the x-line as shown in Fig.~\ref{dissipation_diffusion}(d). The reconnecting component $J_yE_y'$ is also negative but is small compared to the positive $J_zE_z'$ (not shown).

\begin{figure}
\includegraphics[width=8cm]{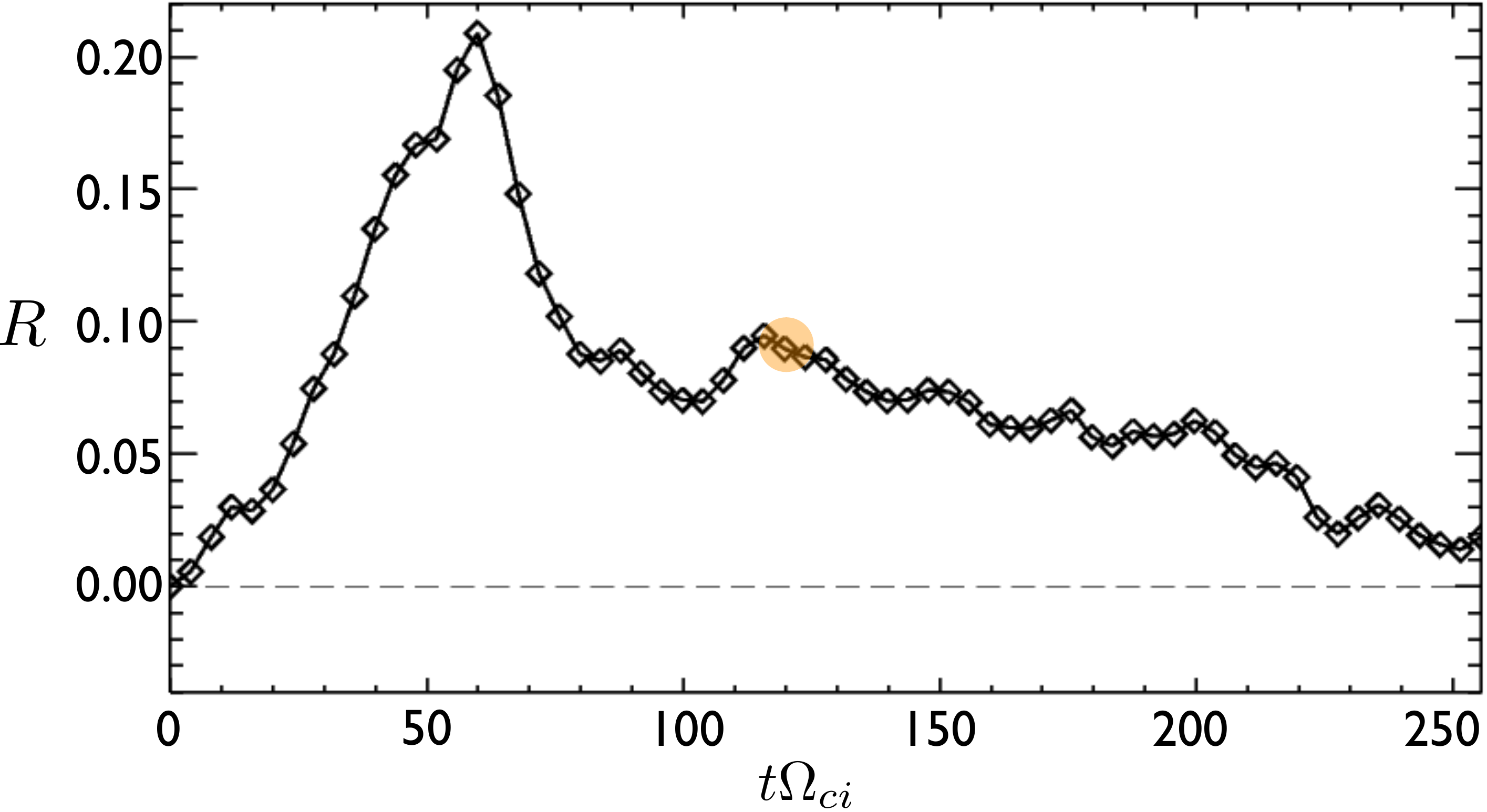} 
\caption {The evolution of normalized reconnection rate. The transparent orange dot at $t\Omega_{ci}=120$ marks the time analyzed in other figures.
}
\label{rate}
\end{figure}

{\it Reconnection rate--} 
A striking feature of this unique case is the value of the reconnection rate, which is largely unaffected by the significant change of the local x-line geometry.
The reconnection rate in our simulation is measured by calculating the change of the in-plane magnetic flux in between the X- and O-points; $R\equiv \left<\partial \Delta \psi/\partial t \right>/(B_0 V_{A0})$ with $\Delta \psi \equiv \mbox{max}(\psi)-\mbox{min}(\psi)$ along the $B_x=0$ trajectory and $\psi$ is the in-plane magnetic flux. As shown in Fig.~\ref{rate}, despite a transient over-shoot at time $t\Omega_{ci}\simeq 60$, the normalized reconnection rate $R$ remains of order of $0.1$ for a considerable duration (in term of ion kinetic time scale), as in a standard case without shear flows \cite{birn01a}. It is imperative to understand why the reconnection rate does not scale as $\delta/L$ \cite{sweet58a,parker57a}, which is larger than unity in this case.

To see how this works, we study the transport of reconnected flux. The electron flow pattern is asymmetric with respect to $z=0$ as in Fig.~\ref{morphology}(d), while the advection of the magnetic flux is symmetric. This difference is allowed by the slippage between plasma and magnetic flux. To take account of the slippage, we generalize the derivation in Ref.~\cite{yhliu16a} to get a proxy of the flux transport velocity in the 2D plane,
\begin{equation}
{\bf U}_{\psi} \equiv{\bf V}_{ep}-({\bf V}_{ep}\cdot \hat{b}_p)\hat{b}_p-c\left(\frac{E_{ey}'}{B_{p}}\right)\hat{b}_{p}\times \hat{y}.
\end{equation}
The subscript ``p'' indicates the in-plane component and the unit vector $\hat{b}_{p}\equiv {\bf B}_{p}/B_{p}$. 
The first two terms quantify the in-plane electron velocity that is perpendicular to the local magnetic field, while the last term represents the slippage velocity between electrons and magnetic flux. This in-plane flux transport velocity satisfies $E_y = -({\bf U}_{\psi}\times {\bf B}_{p})\cdot \hat{y}/c$
by definition. Thus, along the reconnection outflow $E_y = U_{\psi,x}B_z$. The normalized rate is $R\simeq cE_y/B_{x0}V_{A0}\simeq(B_{z}/B_{x0})(U_{\psi,x}/V_{A0})$.

\begin{figure}
\includegraphics[width=8cm]{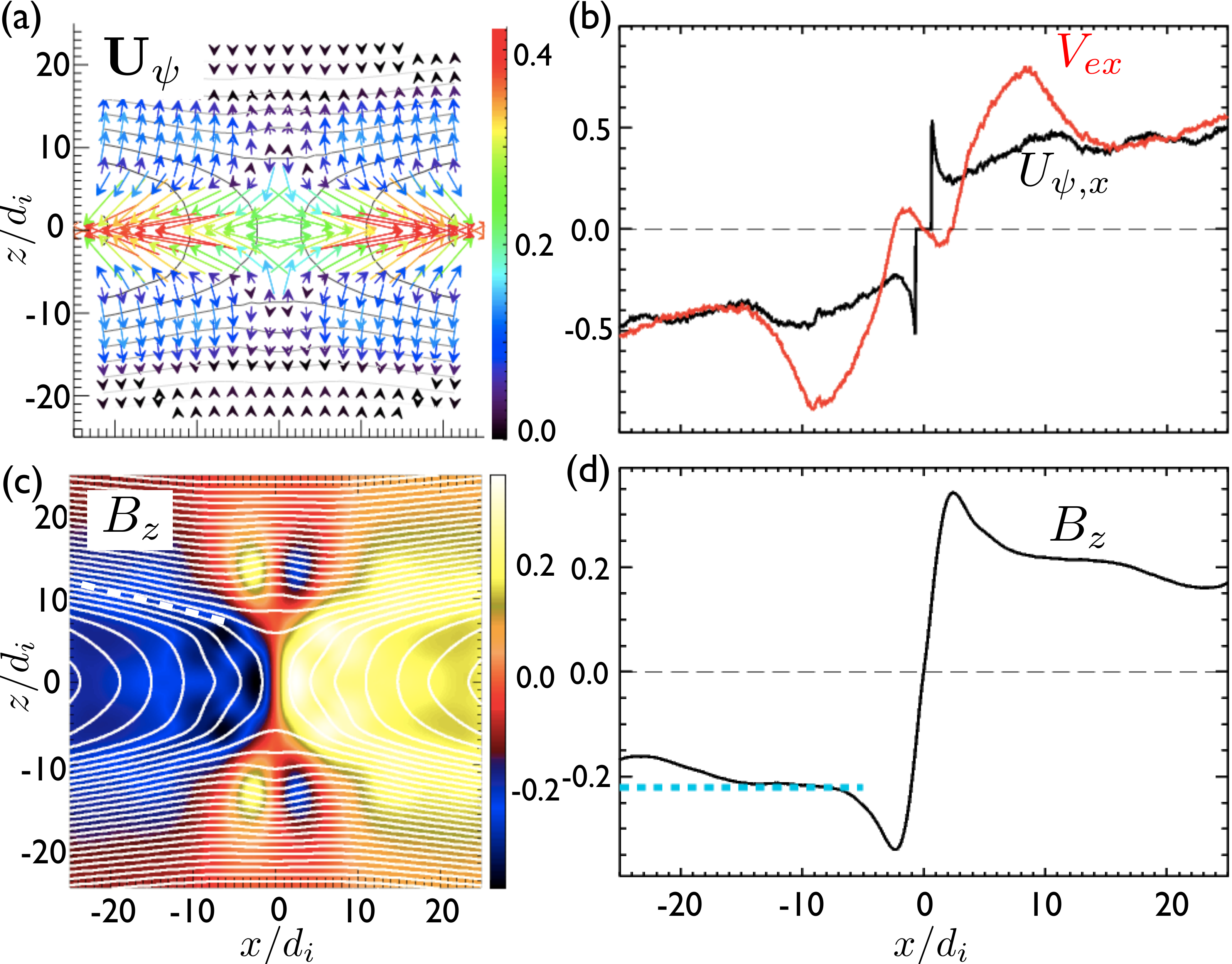} 
\caption {Quantities at time $120/\Omega_{ci}$. (a) Vector plot of flux transport velocity ${\bf U}_{\psi}$. The color represents $|U_{\psi,xz}|\equiv (U_{\psi,x}^2+U_{\psi,z}^2)^{1/2}$. (b) The cut of $U_{\psi,x}$ and $V_{ex}$ at $z=0$. (c) $B_z$ and the contour of $\psi$. The white dashed line marks the opening angle of the upstream magnetic field. (d) The cut of $B_z$ at $z=0$. The cyan horizontal line marks the prediction based on the opening angle in (c). 
}
\label{Upsi}
\end{figure}

This flux transport velocity is symmetric respected to $z=0$ in Fig.~\ref{Upsi}(a), as expected. A cut of $U_{\psi,x}$ at $z=0$ is plotted in Fig.~\ref{Upsi}(b), which reaches a plateau value $\simeq 0.45 V_{A0}$ at $|x| \gtrsim 6 d_i$. The electron velocity $V_{ex}$ is plotted for comparison and it converges to the $U_{\psi,x}$ plateau at $|x| \gtrsim 13 d_i$. 
The reconnected field $B_z$ is shown in Fig.~\ref{Upsi}(c) and a cut at $z=0$ is plotted in Fig.~\ref{Upsi}(d). Although an over-shoot at $|x|\simeq 2d_i$ is necessary to account for the large aspect ratio of the local x-line geometry (i.e., $B_z/B_x\simeq\delta/L$), it plateaus to $\simeq 0.2 B_{x0}$ - a value that we will use to estimate the rate.
The normalized rate is $R\simeq 0.45\times 0.2=0.09$, consistent with the rate in Fig.~\ref{rate} measured using the flux change between the X- and O- points.
 
This analysis shows the following: while the $B_z$ overshoot adjacent to the x-line satisfies the large $\delta/L$ locally, it always approaches a downstream plateau of a lower value. 
The transport of this plateau in $B_z$ by the plateau in $U_{\psi,x}$ better characterizes the quasi-steady reconnection rate. 
This plateau in $B_z$ still satisfies a relation with the opening angle $\theta$ made by the upstream magnetic field \cite{yhliu17a}.
\begin{equation}
\frac{B_{z}}{B_{x0}}\simeq S\frac{1-S^2}{1+S^2}.
\label{Bz}
\end{equation} 
Here $S \equiv \mbox{tan}|\theta| $ is the slope of the upstream magnetic field. The white dashed line along the separatrix in Fig.~\ref{Upsi}(c) measures the opening angle $\simeq 14^\circ$ and its slope is $S\simeq 0.25$. Thus, the expected $B_z\simeq 0.22B_{x0}$ that is comparable to the plateau in $B_z$. Eq.(\ref{Bz}) is obtained by analyzing the force-balance upstream of the diffusion region. The reconnected field $B_z$ is predicted to vanish when $S \rightarrow 1$, which limits the rate when the exhaust opens out. As long as the flux transport speed is Alfv\'enic, the maximum possible rate limited by this upstream constraint (Eq.(\ref{Bz})) is of order 0.1, as also predicted in Ref.~\cite{yhliu17a}. In other words, the upper bound value $\sim O(0.1)$ still applies to the reconnection rate here. (Note that the outflow speed reduction in Ref.~\cite{yhliu17a} is overestimated for this case. The outflow speed does not vanish as predicted for $\delta/L > 1$; the Alfv\'enic outflow continues to be driven by the pair of magnetic kinks as illustrated by the orange dots in Fig.~\ref{model}, working in a fashion similar to Petschek's slow-shock configuration \cite{petschek64a,yhliu12a}).

{\it Summary and discussion--} 
We demonstrated that collisionless magnetic reconnection can proceed even when the current density locally at the x-line has a sign opposite to the initial value, unlike in resistive-MHD. Note that this is different from the coalescence of secondary plasmoids \cite{oka10a}. The field-line dragging by the out-of-plane shear flows provides a distinctly different mechanism that localizes the x-line, but leads to the same reconnection rate $\sim 0.1$; this suggests that the explanation of the fast rate value of order 0.1 in different systems cannot be the result of a ``universal'' localization mechanism.
In particular, the current filamentation tendency of secondary tearing modes does not play any role in regulating the length of the diffusion region in this case because of the flipped current direction. Instead, the reconnection rate can still be explained by the upper bound value provided by the upstream constraint \cite{yhliu17a,cassak17a,yhliu18a}. This strongly localized x-line poses a stringent constraint to any theoretical explanation of the fast rate 0.1.

Caveats need to be kept in mind when applying this result. The embedded current layer eventually becomes unstable at late times ($\sim 250/\Omega_{ci}$) and reconnection rate drops- a phenomenon not studied in this paper.
In the full 3D simulation the interaction of Kelvin-Helmholtz instability (KHI) with reconnection could be important. However, we do not observe a clear flow vortex. The growth of KHI may be reduced by the induced out-of-plane field \cite{XMa16a} and the broadening of the spatial scale of velocity shear. Also, KHI can be suppressed if the perpendicular shear flow is super-fast \cite{miura82a} (Note that $V_{shear} > (V_{A0}^2+v_{th}^2)^{0.5}$ is satisfied in this case). Finally, a similarly embedded current sheet is also observed in simulations with $m_i > m_e$ (not shown). The magnetic geometry becomes asymmetric in the inflow direction because the mass difference between electrons and ions breaks the symmetry of the magnetic flux transport. 
\\

\acknowledgments Y.-H. Liu thanks A. Otto, W. Daughton, B. Rogers and K. -J. Hwang  for
helpful discussions. Y.-H. Liu is supported by NASA grant NNX16AG75G and MMS mission. MH was supported by the Research Council of Norway/CoE under contract 223252/F50, and by NASA's MMS mission. F. Guo and H. Li acknowledge the support by OFES and LANL/LDRD programs. Simulations were performed with LANL institutional computing, NASA Advanced Supercomputing and NERSC Advanced Supercomputing.\\


\end{document}